\documentclass[superscriptaddress,onecolumn,amssymb,amsmath,nobibnotes,
aps,prd,nofootinbib]{revtex4}
\pdfoutput=1
\usepackage{graphicx,subfigure,bm,color,hyperref}
\usepackage{amsfonts}
\usepackage{verbatim}
\usepackage{color}

\begin{document}

		\title{ Note on Morita Inequality for Planar Noncommutative Inverted Oscillator  } 
		
		\author{Praloy Das}
		\email{praloydasdurgapur@gmail.com}
		\affiliation{Physics and Applied Mathematics Unit, Indian Statistical Institute, 203 Barrackpore Trunk Road, Kolkata-700108, India}

		\author{Subir Ghosh}
		\email{subirghosh20@gmail.com}
		\affiliation{Physics and Applied Mathematics Unit, Indian Statistical
			Institute, 203 Barrackpore Trunk Road, Kolkata-700108, India}

		\begin{abstract}
	 A recent conjecture of Morita  predicts a lower bound in temperature $T$ of a chaotic system, $T\geq (\hbar/2\pi)\Lambda$, $\Lambda$ being the Lyapunov exponent, which was demonstrated for a  one dimensional inverse harmonic oscillator. 	In the present work we discuss the robustness of this demonstration in an  extended version of  the above model, where the inverse harmonic oscillator lives  a  in two dimensional noncommutative space. We show that, without noncommutativity, Morita's conjecture survives in an essentially unchanged way in two dimensions. However, if noncommutativity is switched on, the noncommutativity induced correction terms conspire to produce, in classical framework, a purely oscillating non-chaotic system without any exponential growth so that  Lyapunov exponent is not defined. On the other hand, following Morita's analysis, we show that quantum mechanically an effective temperature with noncommutative corrections is generated. Thus Morita's conjecture is not applicable in the noncommutative plane.  A  dimensionless parameter $\sigma =m\alpha\theta^2$, (where $m, \alpha, \theta$ are  the particle mass, coupling strength  with inverse oscillator and the noncommutative parameter respectively) plays a crucial role in our analysis. 
	\end{abstract}
	
\maketitle	

	\section{ Introduction} In a seminal paper Maldacena, Shenker, and  Stanford \cite{mal} argued that a thermodynamic quantum system possesses a sharp upper bound on the rate of growth of chaos. In particular the authors of \cite{mal} conjectured that $\Lambda\leq 2\pi T/\hbar$, $\Lambda$  being the Lyapunov exponent and $T$ the temperature. In a recent interesting work Morita \cite{mor} turned the inequality on its head and came up with an alternative lower bound on temperature:
	\begin{equation}
	T\geq \frac{\hbar}{2\pi} \Lambda,
	\label{0}
	\end{equation}
	with a far reaching consequence that in semi-classical regime a chaotic system (with non-zero Lyapunov exponent) has to have a non-vanishing temperature. An important achievement of \cite{mor} was to demonstrate viability of the above bound in a simple toy   model, a single particle in an inverted oscillator potential. Although, strictly speaking \cite{mor1} the classical model is not chaotic but nevertheless it has a (formally defined) non-zero Lyapunov exponent. In the quantum mechanical framework Morita \cite{mor} showed that the tunneling phenomena induces an effective non-zero temperature resulting in a net (analogue) Hawking flux. Interestingly enough, the effective temperature generated in the quantum framework saturates the bound (\ref{0}). This demonstration was possible primarily because the classical Lyapunov exponent  is exactly and analytically computable for the simple system considered in \cite{mor,mor1}.
	
	The question that naturally comes to mind is how robust is this demonstration of validity of the inequality (\ref{0}) that is whether it will survive in a more complicated (but still toy!) model. Towards this end we consider a (spatially) Non-Commutative (NC) extension of this model which also brings in another extension in dimension: a two dimensional non-relativistic system is essential to have a non-trivial non-commutativity. Subsequently we rederive results analogous to that of Morita \cite{mor,mor1}. In recent times, extending models, both non-relativistic as well as relativistic, to NC space and spacetime, has paid rich dividends (for reviews see for example \cite{ncrev}) since it introduces a dimensional  scale via the NC parameter $\theta_{ij}$. Different structures of NC generalization brings in new interactions in the resulting theories and has been effective in ushering in possible quantum gravity corrections in a phenomenological perspective.
	
	The paper is organized as follows. In Section II the NC inverted oscillator in two dimensional space is introduced, along with the Darboux-type of canonical phase space that we subsequently exploit. In Section III we discuss the above model in quantum mechanical framework where NC corrections in the effective temperature are computed. Section IV is devoted to a classical analysis of the NC inverted oscillator. In Section V we study commutative version of the above model. Section VI contains a summary of the work presented together with future directions. In Appendices 1 and 2 some detailed results are appended.
	
	\section{    Inverted oscillator in noncommttative plane} The planar Hamiltonian of the particle in inverted oscillator potential is given by
	\begin{equation}
	H=\frac{m}{2}\dot{x_i}\dot{x_i}-\frac{\alpha}{2}x_ix_i~,~~i=1,2
	\label{1}
	\end{equation}	
	along with the NC-extended classical Poisson bracket structure,
	\begin{equation}
	\{x_i,p_j\}=\delta_{ij},~~\{p_i,p_j\}=0,~~\{x_i,x_j\}=\theta_{ij}=\theta\epsilon_{ij};~~\epsilon_{12}=1.
	\label{2}
	\end{equation}	
	It is straightforward to obtain the NC equation of motion (in terms of non-canonical variables)
	\begin{equation}
	\dot x_i=\{x_i,H\}=\frac{1}{m}p_i-\alpha\theta\epsilon_{ij}x_j,~\dot p_i=\{p_i,H\}=\alpha x_i $$$$ \ddot x_i=\frac{\alpha}{m}x_i-\alpha\theta \epsilon_{ij}\dot x_j ,
	\label{3}
	\end{equation}
the last equation (\ref{3}) being a modified version of Newton's law. 	In a classical scenario non-canonical (instead of non-commutative) degrees of freedom  would be a better nomenclature but we stick to the conventional one.	We use {\it time, mass, distance} as the fundamental dimensional units, in terms which $[x_i]=length$ and the constant parameters have $[\alpha ]=(mass)/(time)^2, ~[\theta ]=(time)/(mass)$. It is conventional and more convenient to use the canonical variables $X_i,P_i$ obeying  $\{X_i,X_j\}=\{P_i,P_j\}=0;~\{X_i,P_j\}=\delta_{ij}$ such that $x_i,p_i$ can be mapped to $X_i,P_i$ ($\sim$ Darboux map):
	\begin{equation}
	x_1=X_1-\frac{\theta}{2}P_2,~~x_2=X_2+\frac{\theta}{2}P_1,~~p_1=P_1,~~p_2=P_2 .
	\label{4}
	\end{equation}	
	The Hamiltonian is rewritten as
	\begin{equation}
	H=\frac{\bar a}{2}P_iP_i-\frac{\alpha}{2}X_iX_i+\frac{\theta\alpha }{2}\epsilon_{ij}X_iP_j; ~~\bar a=\frac{1}{m}-\frac{\theta^2\alpha}{4},
	\label{5}
	\end{equation}	
	where the NC modification has induced a constant magnetic field like effect \cite{sg1}.

	\section{ NC effect on quantum dynamics} In the quantum version we use the polar coordinates  $X_1=\rho cos\phi, X_2=\rho sin\phi $ in order to exploit the rotational symmetry of the problem where $\phi$ becomes cyclic and the problem reduces to a one dimensional one with the Hamiltonian   
	\begin{equation}
	H=\big\{-\frac{\hbar^2\bar a}{2}(\partial_\rho ^2 +\frac{1}{\rho }\partial_\rho +\frac{1}{\rho^2}\partial_\phi ^2)-\frac{\alpha}{2}\rho^2+\frac{i\hbar \theta\alpha}{2}\partial_\phi \big\}
	\label{9}
	\end{equation}
	to be used in  time independent Schrodinger equation $H\psi = E\psi $.	For $\psi$ of the form 
	$\psi(\rho,\phi)=F(\rho)exp(iL\phi)$ the equation for $F(\rho)$
	\begin{equation}
	\partial_\rho ^2 F+\frac{1}{\rho}\partial_\rho F-\frac{L^2}{\rho^2}F+\frac{\alpha}{\hbar^2 \bar a}\rho^2 F +\frac{2}{\hbar^2 \bar a}(\frac{\hbar\theta\alpha L}{2}+E)F=0
	\label{10}
	\end{equation}	
	re-expressed in terms of dimensionless variables $z_{NC},  E_{NC}$  \cite{mor} 
	\begin{equation}
	z_{NC} =(\frac{4\alpha }{\hbar^2\bar a})^{1/4}\rho,~- E_{NC}=\sqrt{\frac{1}{\hbar^2\alpha \bar a}}(E+\frac{\hbar\theta\alpha L}{2}). 
	\label{11}
	\end{equation}	
	turns out to be,
	\begin{equation}
	\partial_{z_{NC}} ^2 F+\frac{1}{z_{NC}}\partial_{z_{NC}} F-\frac{l^2}{z_{NC}^2}F+(\frac{z_{NC}^2}{4}  - E_{NC})F=0.
	\label{12}
	\end{equation}	
	Comparing with the analogous equation A3 of \cite{mor1} we find that there is a qualitative change due to additional $1/z_{NC}$-term and the $1/z_{NC}^2$ centrifugal term apart from NC-corrections of the constant parameters.
	\begin{figure}
		\begin{center}
			\includegraphics[width=0.6\textwidth]{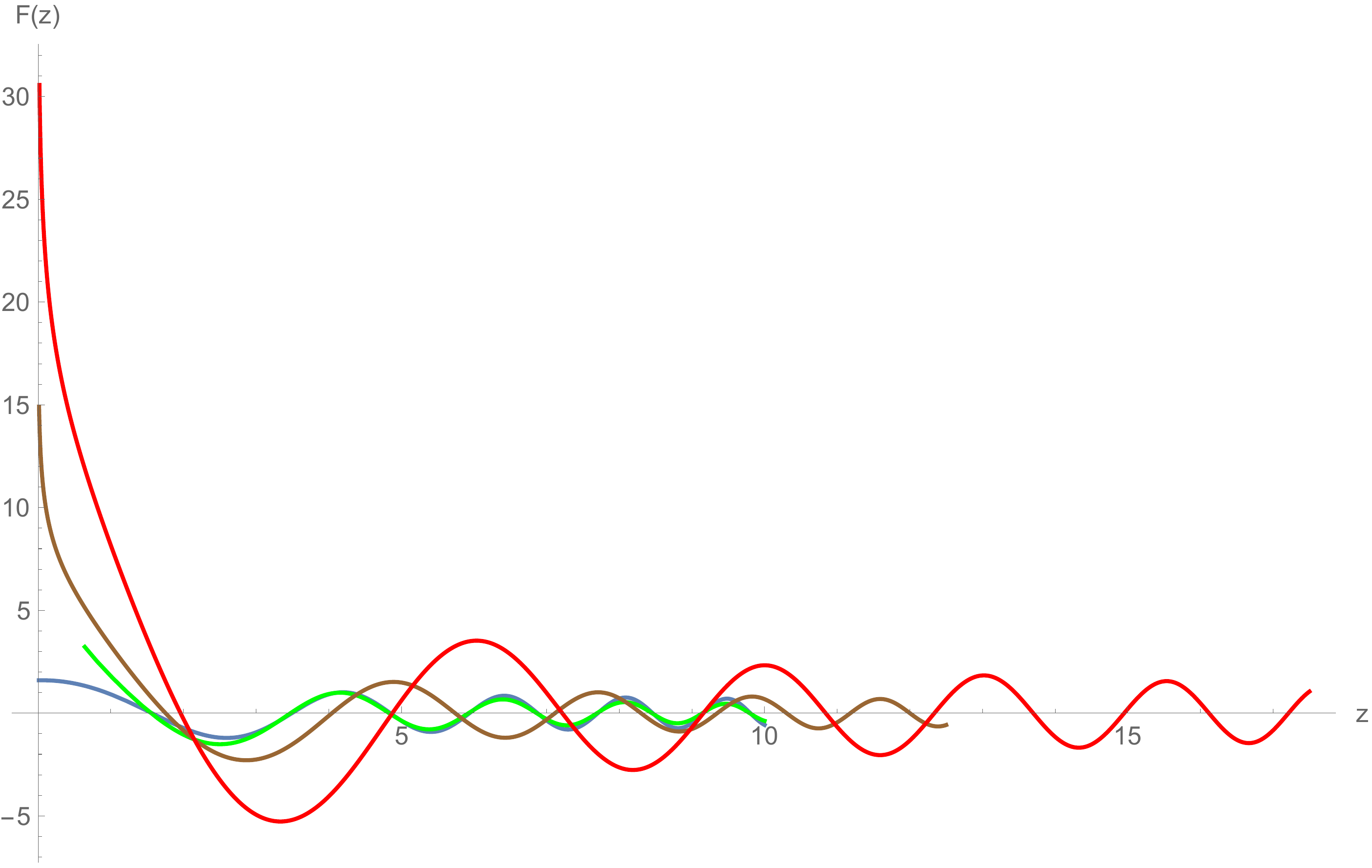}
			\caption{Real part of solution of (\ref{12}) (hypergeometric function) for $\sigma=0$, green curve; $\sigma=4$, brown curve and $\sigma=12$, red curve and parabolic cylinder function of \cite{mor,mor1}, blue curve, are plotted for $L=0$ in each case. }
			\label{Re(M=0)}
		\end{center}
	\end{figure}
	
	\begin{figure}
		\begin{center}
			\includegraphics[width=0.6\textwidth]{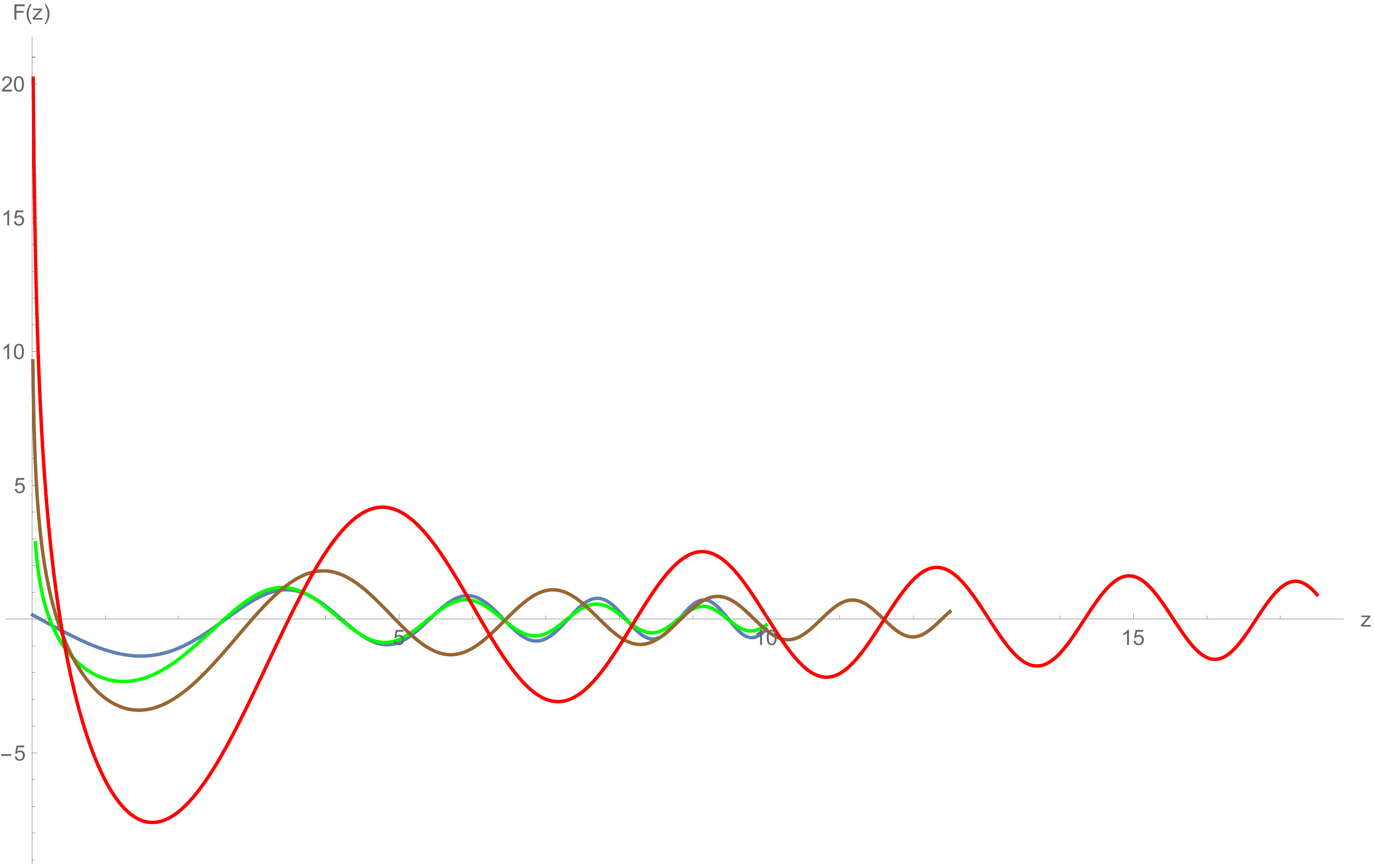}
			\caption{Imaginary part of solution of (\ref{12}) (hypergeometric function) for $\sigma=0$, green curve; $\sigma=4$, brown curve and $\sigma=12$, red curve and parabolic cylinder function of \cite{mor,mor1},blue curve, are plotted for $L=0$ in each case. }
			\label{Im(M=0)}
		\end{center}
	\end{figure}
	
	 We are interested in the large $z_{NC}$ behavior \cite{mor,mor1} in which case the simplest and crudest option is to drop these new terms  from  (\ref{12}) thus yielding  
	\begin{equation}
	\partial_{z_{NC}} ^2 F+(\frac{z_{NC}^2}{4}  - E_{NC})F=0.
	\label{12a}
	\end{equation}
	Now (\ref{12a}) is structurally identical to the equation considered in \cite{mor1} and all we need to do is to replace $E$ by $ E_{NC}$ containing the NC and non-zero $L$ corrections. Hence for large $z_{NC}$ 
	\begin{equation}
	F(z_{NC})\approx C(1)\Gamma_1 (L)e^{i(\frac{z_{NC}^2}{4}- E_{NC}~\log\,z_{NC} )} $$$$+C(2)\Gamma_2 (L)e^{-i(\frac{z_{NC}^2}{4}- E_{NC}~\log\, z_{NC} )}
	\label{14}
	\end{equation}	
	where $\Gamma_i(L)$ are $L$-dependent numerical constants and   $ z_{NC}, E_{NC}$ are explicitly given by
	\begin{equation}
	z_{NC}=(\frac{4\alpha }{\hbar^2 a})^{1/4}(1-\frac{\sigma}{4})^{-\frac{1}{4}} \approx  z_M(1+\frac{\sigma}{16})+O(\sigma^2), $$$$
	E_{NC}\approx -E_M(1+\frac{\sigma}{8})-{\sqrt{\sigma}}L +O(\sigma^{3/2}).
	\label{15}  
	\end{equation}
	In the above the subscript $M$ refers to Morita's parameterization \cite{mor,mor1} and finally we have kept up to $O(\theta^2)$ terms. Once again, following \cite{mor1} we easily obtain the tunneling probability
	\begin{equation}
	P_T(E)=\frac{1}{{  e^{-\frac{2\pi E}{\hbar}\sqrt{\frac{m}{\alpha}}(1-\frac{\sigma}{8})}~e^{\pi L \sqrt{\sigma}}}+1}   =  \frac{1}{{  e^{-\frac{2\pi E_M}{\hbar}(1-\frac{\sigma}{8})}~e^{\pi L \sqrt{\sigma}}}+1}
	\label{18}
	\end{equation}
	which for $L=0$ provides the temperature
	\begin{equation}
	T_{NC}= \frac{\hbar}{2\pi} {\sqrt \frac{\alpha}{m}}(1-\frac{\sigma}{8})^{-1}\approx T_M (1+\frac{\sigma}{8}).
	\label{19}
	\end{equation}
	Thus an NC correction appears through $\sigma$ in the effective temperature. The effect of non-zero $L$ can be read off from (\ref{19}). 
		\begin{figure}
		\begin{center}
			\includegraphics[width=0.6\textwidth]{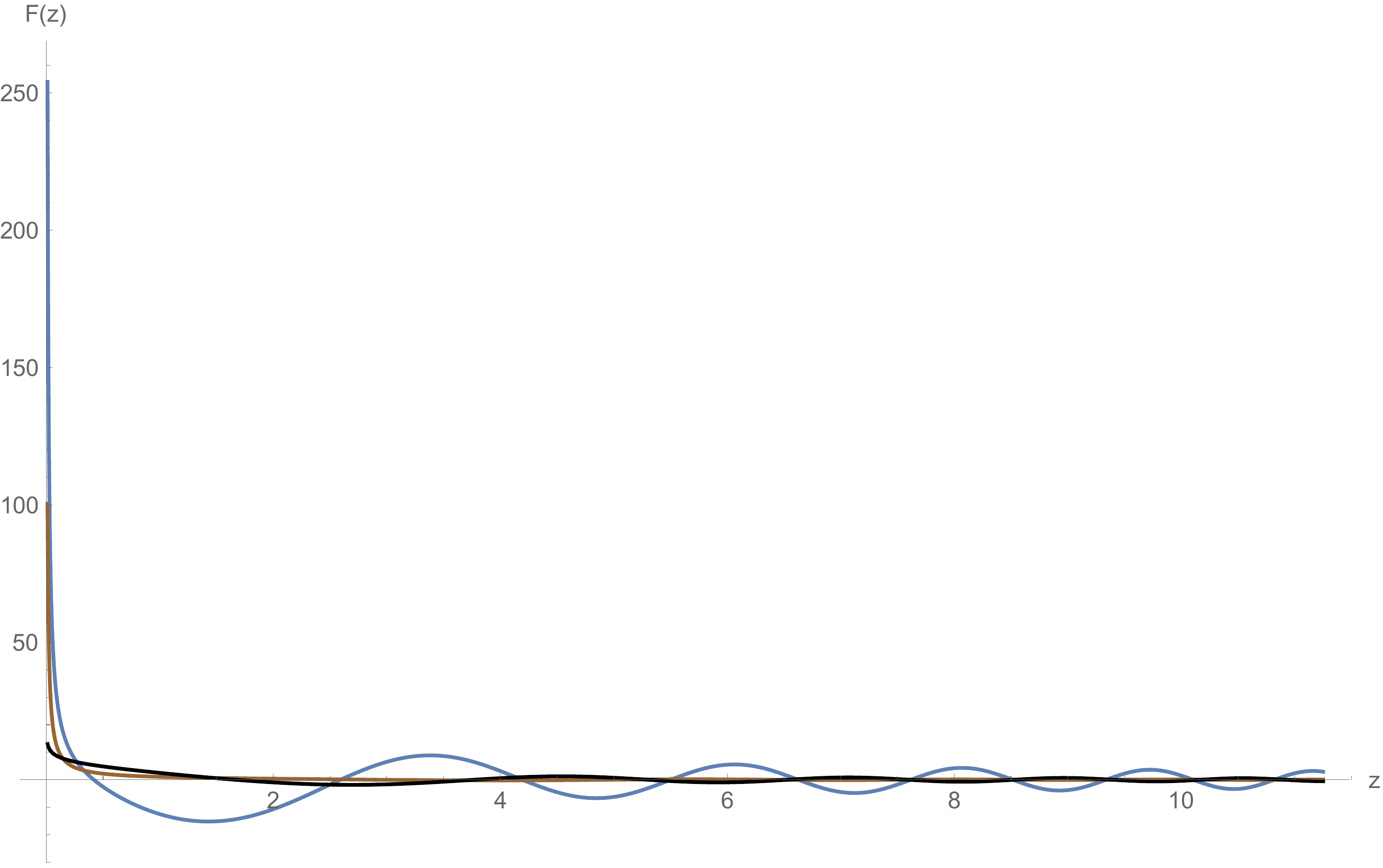}
			\caption{Real part of solution of (\ref{12}) (hypergeometric function) for $\sigma=2$, and different values of $L=0$, black curve; $L=1$, blue curve and $L=-1$, brown  curve are plotted. }
			\label{re(M=+-1,sigma=2)}
		\end{center}
	\end{figure}
	
	In Appendix 1, in (a7),  we have provided full solution of (\ref{12}) which comprises of a combination of two linearly independent solutions given by hypergeometric and Laugurre functions. The solutions of the one dimensional commutative model \cite{mor1} appear as parabolic cylinder functions. We have picked the Hypergeometric  function in our case since in the commutative limit it is closer to the parabolic cylinder solutions. In Figure 1 and Figure 2 we have plotted respectively real and imaginary parts of the Hypergeometric  function for $L=0$ and different values of $\sigma =0,4,12$. we have also included profiles of the real and imaginary parts of the parabolic cylinder function for the sake of comparison.
	
	In Figure 3 the profiles of real parts of the solution of (\ref{12}) for fixed $\sigma =2$ and different values of $L=0,\pm 1$ are drawn. A common feature in all the NC and $L\neq 0$ graphs is that the peak at $z=0$ is much higher than the one dimensional commutative model of \cite{mor,mor1} but the solutions die out for large $z$ in a similar fashion, which is not surprising.

\section	{ NC effect on classical dynamics} Using (\ref{5}) the equations of motion
		\begin{equation}
	\dot X_i=\{X_i,H\} = (\frac{1}{m}-\frac{\alpha\theta^2}{4}) P_i-\frac{\alpha\theta}{2}\epsilon_{ij}X_j, 
		~~\dot P_i=\{
	P_i,H\} =\alpha X_i-\frac{\alpha\theta}{2}\epsilon_{ij}P_j
	\label{6a}
	\end{equation}
			 can be decoupled for $X_i$ to yield a fourth order equation of motion,
	\begin{equation}
	X_i^{(.4)}+(B^2-2A)X_i^{(.2)}+A^2X_i =0
	\label{6}
	\end{equation}	
	where $X^{(.n)}=(d^n X)/(dt^n)$ and $A=\bar a \alpha +\frac{\theta^2\alpha^2}{4}=1/m ,~B=-\theta \alpha^2 $. Incidentally this is reminiscent of the Pais-Uhlenbeck oscillator dynamics \cite{pai,pais} although we will not go in to that topic in the present work.
	
	The four linearly independent solutions are
	\begin{equation}
	X_i=C^{(\mu)}exp \large (\pm {\sqrt {A-\frac{B^2}{2}\pm {\sqrt {B^4-4AB^2}}}}\large ),
	\label{7}
	\end{equation}	
	where each  independent solution  has  integration constant $C^{(\mu)},\mu=1,2,3,4$ and the two $\pm$ signatures operate independently. Defining a dimensionless parameter $\sigma =m\alpha\theta^2$ the exponents of the solutions appear as
	\begin{equation}
	\omega_{NC} (\mu)= \pm {\sqrt {\frac{\alpha}{m} (1-\frac{\sigma}{2}\pm \frac{1}{2}  {\sqrt {\sigma^2-4\sigma}}})} = \pm \omega_M  {\sqrt  {(1-\frac{\sigma}{2}\pm \frac{1}{2}  {\sqrt {\sigma^2-4\sigma}}})}.
	\label{8}
	\end{equation}
	In (\ref{8}) subscripts $NC$ and $M$ stand for noncommutative and Morita respectively. 	For $\sigma =0$ the result of Morita \cite{mor} is recovered with Lyapunov exponent $\omega_M (\sigma =0)=\sqrt{\alpha /m}$. Note that although $\theta$ is expected to be small $\sigma$ can be large with proper tuning of $\alpha,m$.

		The general solutions can be expressed as,
	\begin{eqnarray}
	X_1=A_1 e^{(\omega_1 t)}+A_2 e^{-(\omega_1 t)}+A_3 e^{(\omega_3 t)}+A_4 e^{-(\omega_3 t)} \nonumber \\
	X_2=B_1 e^{(\omega_1 t)}+B_2 e^{-(\omega_1 t)}+B_3 e^{(\omega_3 t)}+B_4 e^{-(\omega_3 t)}
	\label{a3}
	\end{eqnarray}	
	where
	$$\omega_{1(2)}= +(-) \omega_M  {\sqrt  {(1-\frac{\sigma}{2}+ \frac{1}{2}  {\sqrt {\sigma^2-4\sigma}}})};~\omega_{3(4)}= +(-) \omega_M  {\sqrt  {(1-\frac{\sigma}{2}- \frac{1}{2}  {\sqrt {\sigma^2-4\sigma}}})} $$
 are the four possible values of $\omega_{NC}$ in (\ref{8}).	Substituting the solutions in (\ref{6a}) the integration constants $A_i,B_i$ are restricted to the following relations:
	\begin{eqnarray}
	\frac{A_1}{B_1}=-\frac{A_2}{B_2}=\frac{\omega_1^2-\alpha/m}{\alpha\theta\omega_1}\equiv\Lambda_1;~~
	\frac{A_3}{B_3}=-\frac{A_4}{B_4}=\frac{\omega_3^2-\alpha/m}{\alpha\theta\omega_3}\equiv\Lambda_2.
	\label{a5}
	\end{eqnarray}
	\begin{figure}
		\begin{center}
			\includegraphics[width=0.6\textwidth]{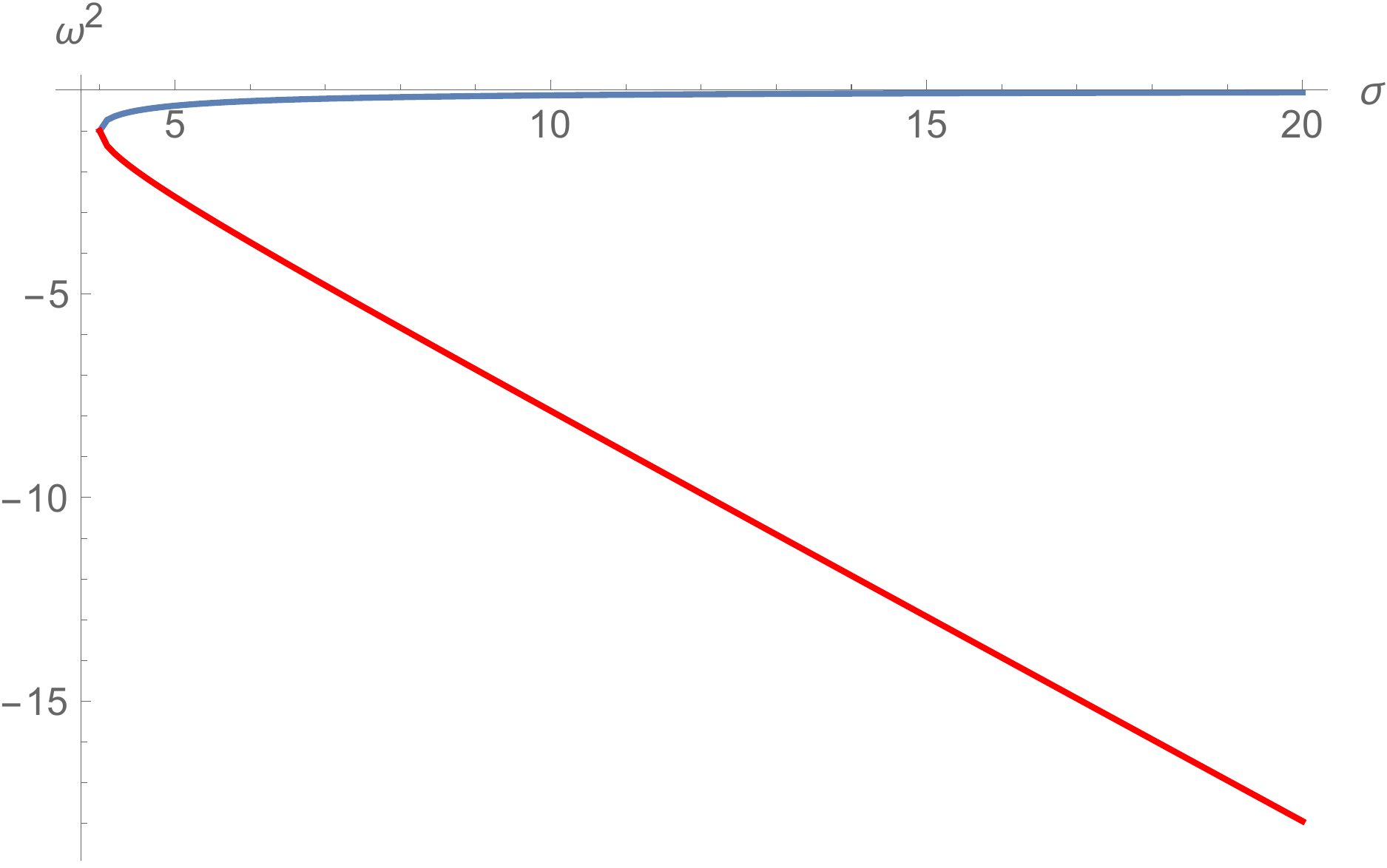}
			\caption{Plot of $\omega^2$ against $\sigma$ for the two roots, $\omega_1^2$, blue line, and $\omega_3^2$, red line, showing that $\omega^2$ is always negative for the allowed range of $\sigma$. }
			\label{linear}
		\end{center}
	\end{figure}
	Hence, the solutions appear as
	\begin{eqnarray}
	X_1=\Lambda_1 A(e^{\omega_1 t}-B e^{-\omega_1 t})+\Lambda_2 C(e^{\omega_3 t}-D e^{-\omega_3 t})~~ , \nonumber \\
	X_2= A(e^{\omega_1 t}+B e^{-\omega_1 t})+ C(e^{\omega_3 t}+D e^{-\omega_3 t}).
	\label{b6}
	\end{eqnarray}
	Using (\ref{4}) in (\ref{5}) the Hamiltonian is written completely in terms of $X_i,\dot X_i$ 
		\begin{eqnarray}
	H=\frac{1}{2a}(\dot{X_1}^2+\dot{X_2}^2)-\frac{\alpha}{2}X^2-\frac{\alpha\theta}{4a}(X_1\dot{X_2}-X_2\dot{X_1}).
	\label{a7}
	\end{eqnarray}
Finally using the above explicit forms of $X_i$ given in (\ref{b6}) in the above equation, we find expression for the energy,
\begin{eqnarray}
H&=& A^2 B[(\Lambda_1^2-1)(\frac{\omega_1^2}{a}+\alpha)+\frac{\alpha\theta}{a}\Lambda_1\omega_1]+C^2 D[(\Lambda_2^2-1)(\frac{\omega_3^2}{a}+\alpha)+\frac{\alpha\theta}{a}\Lambda_2\omega_3] \nonumber \\
&& +2\alpha AC[e^{(\omega_1+\omega_3)t}+BDe^{-(\omega_1+\omega_3)t}] .
\label{ab8}
\end{eqnarray}
Notice that because of the mixing of $\omega_1$ and $\omega_3$ in $X_i$ in (\ref{b6}) there appears an unwanted time dependence in $H$ from the last two terms of (\ref{ab8}). To avoid that we choose $X_i$ to be of the form,
	\begin{eqnarray}
X_1=\Lambda_1 A(e^{\omega_1 t}-B e^{-\omega_1 t})~~ , ~~
X_2= A(e^{\omega_1 t}+B e^{-\omega_1 t})
\label{a6}
\end{eqnarray}
with energy 
\begin{eqnarray} 
H=-A^2 B [\frac{1}{a}(\omega_1^2+\frac{\alpha}{m})+2\alpha ] =-2\alpha A^2 B[3+\frac{2\sqrt{\sigma^2-4\sigma}}{(4-\sigma)}]
\label{a10}
\end{eqnarray}
or 
	\begin{eqnarray}
X_1=\Lambda_2 C(e^{\omega_3 t}-D e^{-\omega_3 t})~~ , ~~
X_2=  C(e^{\omega_3 t}+D e^{-\omega_3 t})
\label{a6}
\end{eqnarray}
with energy
\begin{eqnarray} 
H=-C^2 D [\frac{1}{a}(\omega_3^2+\frac{\alpha}{m})+2\alpha ] =-2\alpha C^2 D[3-\frac{2\sqrt{\sigma^2-4\sigma}}{(4-\sigma)}].
\label{a10}
\end{eqnarray}
Furthermore, requiring $H$ to be real one has to impose the condition $\sigma^2-4\sigma \geq 0$ so that $\sigma=0$ (commutative case) or $\sigma\ge 4$. 	Now comes the surprising result. For $\sigma\geq 4$ both $(\omega_1)^2=(\Lambda^1_{NC})^2$ and $(\omega_3)^2=(\Lambda^3_{NC})^2$ turn out to be {\it {always negative}} where $\Lambda_{NC}$ notation is used to identify them with NC corrected Lyapunov exponents. This is shown in  Figure 4 where we have plotted $\omega^2$ against $\sigma$. The two colored lines correspond to the two roots, $\omega_1^2$ for blue line and $\omega_3^2$ for red line.  This in turn means that the $\Lambda^1_{NC}$ and $\Lambda^3_{NC}$ are purely imaginary thereby restricting the solutions to be purely oscillatory without any exponentially divergent chaotic behavior. Since there are no Lyapunov exponents it is not possible to verify Morita's conjecture.
	Thus, (even though in (\ref{19}) $\sigma$-correction is qualitatively similar to (\ref{8}) actually the latter is purely imaginary), we conclude that the planar noncommutative inverted oscillator model is not appropriate to discuss the  Morita conjecture \cite{mor, mor1}.
	
\section{Planar commutative inverted oscillator}	In the classical planar commutative case with $\theta=0~\rightarrow \sigma =0$ (considered by us) the frequencies are same as the case considered by Morita, i.e. $\omega_{NC}=\omega_M$ with same Lyapunov exponent. On the other hand, as we have discussed earlier in Section 3, for large $z$ the equation (\ref{12a}) for $\sigma=0$ reduces to the one considered in \cite{mor,mor1}, and hence the Morita conjecture will be verified in the same way. In Appendix 1	 The solutions (computed with Mathematica package) 
of both (\ref{12}) and (\ref{12a}), (the latter for the commutative case $\sigma =0$), are given. Note that, as observed in \cite{mor,mor1}, for large $\rho \sim z_{NC}$, both the solutions oscillate as $G\,exp(i\chi)$ with identical form of $\chi \approx z_{NC}^2/4 -  E_{NC}\, \log\,z_{NC} $ but differing in $G$ which varies as $1/\sqrt{z_{NC}}$ and  $1/z_{NC}$   in (\ref{12a}) and (\ref{12}) respectively. Since the large distance behavior is responsible for the transition and reflection rates we can safely assume that we will reach   similar conclusion as (\ref{18}, \ref{19}).

It is worthwhile to mention one point in our classical analysis that is discussed in detail in Appendix 2. Recall that we have used polar coordinates in quantum analysis and Cartesian coordinates in the classical analysis and so it seems natural to relate the quantum and classical results for $L=0$ case only. It would have been more convenient to study the classical case in polar coordinates as well but recasting the noncommutative phase space algebra in polar form (and the subsequent Darboux map) is quite difficult and we have not been able to achieve a satisfactory solution (see Appendix 2 for details). This has forced us to do the classical analysis in cartesian coordinates.

		\section{ Discussion}  The present work is the first attempt to generalize the recent important conjecture of Morita \cite{mor,mor1} that proposes a lower bound in the temperature of a chaotic semi-classical system. The demonstration in \cite{mor,mor1} exploits the simplest possible toy model - one dimensional system consisting of a particle in inverted oscillator potential. In this case the classically computed Lyapunov exponent yields the bound (via Morita's relation) that saturates the effective temperature derived from quantum analysis of the model. We have extended the model in two directions by considering the same model in   two  dimensional noncommutative
	space.
	
	In the two dimensional version in canonical space we find that Morita's conjecture is satisfied, that is, even though in the quantum analysis, two dimensional configuration space generates additional non-trivial terms in the Hamiltonian, in the large distance limit these can be neglected thus leading to no qualitative change in the results and conclusion. On the other hand, quite surprisingly (and somewhat disappointingly!) we find the the noncommutative extension does not allow the application of Morita's inequality because, in the classical analysis, the noncommutive correction terms conspire to render the system purely oscillatory, hence non-chaotic. Although, as we have shown, the quantum analysis does yield an effective temperature with noncommutative corrections, there is no way to relate it to the Lyapunov exponent in its classical version simply because the latter does not exist. 
	
	Furthermore, as our analysis has revealed,  setting up a classical noncommutative coordinate algebra in polar form and deriving the resulting canonical solution (by way of a local Darboux map) is quite involved and unsolved.
	
	As future extension of the present work an immediate  task is to compare the above mentioned results more rigorously, in particular by formulating the noncommutative geometry in polar coordinates.\\ One option is to incorporate other, more involved forms of noncommutativity (such as the algebra compatible to Generalized Uncertainty Principle or the Snyder form of noncommutativity).\\ Another interesting area of research would be the fluid model. In Morita's paper \cite{mor,mor1} it has been shown that in Fermi fluid, consisting of non-relativistic fermions obeying the canonical fluid equations (continuity and Euler equations with characteristic pressure term), acoustic Hawking radiation appears above a critical fluid velocity. We have constructed noncommutative extensions  of ideal fluid model in a series of earlier works \cite{sgfluid, ncfl} and it would indeed be worthwhile to test Morita's conjecture in such system. \\

	{\bf Acknowledgments:} It is a pleasure to thank Takeshi Morita for many helpful correspondences. The work of P.D. is supported by INSPIRE, DST, India. \\
	
	{\bf Appendix 1:} The solution  for (\ref{12}) using Mathematica is given by
\begin{eqnarray}
F(z_{NC}) & = & 2^{(1 + L)/2}e^{-i z_{NC}^2/4} (z_{NC})^L \left\{ C(1)\,F[ (1 + L - i  E_{NC})\frac{1}{2}, (1 + L), \frac{i z_{NC}^2}{2}]\right. \nonumber \\ & + & \left. C(2)L_G[ (-1 - L + i  E_{NC})\frac{1}{2},  L, \frac{i z_{NC}^2}{2}] \right \},
\label{a7}
\end{eqnarray}
where $F$ and $L_G$ are respectively the hypergeometric and Laugurre functions.
For large $z$ this reduces to
\begin{eqnarray}
F(z_{NC}) &\approx &   C(1)e^{-iz_{NC}^2/4}z_{NC}^{i E_{NC}}\left\{\frac{(-1)^{(7-L+i E_{NC})/4}\,2^{(1+L-i E_{NC}/2)}}{z_{NC}}+O(1/z_{NC}^2)\right\}\nonumber \\ & &
+\,C(2)e^{-iz_{NC}^2/4}z_{NC}^{i E_{NC}}\left\{(\frac{(-1)^{1/4}\,2^{(1+L-i E_{NC}/2)}\,e^{(iL+ E_{NC})\pi/4}}{\Gamma[(1-L+i\bar E)/2]\,z_{NC}})+O(1/z_{NC}^2)\right\}\nonumber \\ & &
+\,C(2)e^{iz_{NC}^2/4}z_{NC}^{-i E_{NC}}\left\{\frac{(-1)^{(7-L)/4} \,e^{\pi E_{NC}/4}\,cos[(L-i E_{NC})\pi/2]}{\pi z_{NC} (2^{-(1+L+i E_{NC}/2)})}\,
\Gamma[(1+L+i E_{NC})/2]  +O(1/z_{NC}^2) \right\}. \nonumber \\
\label{a8}
\end{eqnarray}
On the other hand the result of Morita \cite{mor1}, solution of (\ref{12a}) with $Q$ denoting the parameter $ E_{NC}$, is given by 
\begin{equation}
F_M(z)=C(1)D[(i+2Q)i/2,(-1)^{1/4}z ] +C(2)D[-(-i+2Q)i/2,(-1)^{3/4}z ]
\label{a9}
\end{equation}
where $D$-functions are parabolic cylinder functions, 
with the large $z$ approximation given by
\begin{eqnarray}
F_M(z) &\approx & e^{iz^2/4}z^{-iQ}\left\{-C(2)\sqrt{\frac{1}{z}}(-1)^{5/8}e^{3\pi Q/4} +O(1/z)^{3/2}\right\}  +\, e^{-iz^2/4}z^{iQ}\left\{-C(1)\sqrt{\frac{1}{z}} (-1)^{7/8}e^{-\pi Q/4} \right\} \nonumber \\& & +\, e^{-iz^2/4}z^{iQ}\left\{C(2)\sqrt{\frac{1}{z}} \frac{1}{\Gamma[1/2+iQ]}(-1)^{1/8}e^{\pi Q/4}\sqrt{2\pi} 
  + O(1/z^{3/2})  \right\}.
\label{a10}
\end{eqnarray}
	 
	{\bf Appendix 2:} It is worthwhile to mention one point of our analysis. Recall that we made the comparison between  classical and  semi-classical NC corrected results, treating the latter in polar coordinates and the former in Cartesian coordinates. Hence it seems natural to consider the $L=0$ case in the quantum scenario for comparing. It is pertinent to ask why we did not use polar coordinates in the classical case. We now reveal a surprising and subtle facet that it is quite complicated to treat a simple {\it {two-dimensional noncommutative system in polar coordinates}}. Let us try to generate these noncommutative space brackets dynamically. In Dirac's \cite{dir} Hamiltonian formulation of constraint analysis, we start from the Lagrangian (in Cartesian coordinates)
	\begin{equation}
	L=\frac{1}{2\theta}(x_1\dot x_2-x_2\dot x_1)=\frac{1}{2\theta}\epsilon_{ij}x_i\dot x_j	
	\label{a1}
	\end{equation}
	with conventional Poisson brackets 
	$ \{x_i,x_j\}= \{p_i,p_j\}=0,~	 \{x_i,p_j\}=\delta_{ij}$. The noncommuting (in the sense of Poisson brackets) pair of Second Class constraints (in Dirac's classification)
	\begin{equation}
	\chi_i\equiv p_i+\frac{1}{2\theta}\epsilon_{ij}x_j ;~~\{\chi_j,\chi_j\}=-\theta \epsilon_{ij}\equiv \chi_i{ij}.	
	\label{a2}
	\end{equation}
	Consistent implementation of the constraints is given by the Dirac bracket, $\{A,B\}_{D}$, which for two generic variables $A,B$, is defined by,
	\begin{equation}
	\{A,B\}_{D}=\{A,B\}-\{A,\chi_i\}\{\chi_i,\chi_j\}^{-1}\{\chi_j,B\}.
	\label{d}
	\end{equation}
	In the present case the Dirac brackets  (with the subscript $D$ omitted) are given by 
	\begin{equation}
	\{x_i,x_j\}=-\theta\epsilon_{ij},~	\{x_i,p_j\}=\frac{1}{2}\delta_{ij},~	 \{p_i,p_j\}=-\frac{1}{4\theta}\epsilon_{ij}.
	\label{a3}
	\end{equation}
	However recall that instead of the above brackets we have used (\ref{2}), i.e. $	\{x_i,x_j\}=-\theta\epsilon_{ij},~	\{x_i,p_j\}=\delta_{ij},~	 \{p_i,p_j\}=0$. We are allowed to do it because this deformation is not operatorial in nature and algebraically does not clash with the Jacobi identities among $x_i,p_i$. (Dirac brackets satisfy Jacobi identity.)

	Let us now come to polar coordinates $\rho,\phi $ defined by  	$x_i=\rho cos\phi,x_2=\rho sin\phi$.  Indeed, deriving the NC brackets for polar variables, in a purely algebraic way, starting from Cartesian one, is quite involved (see eg. \cite{ncpolar}). However generating them dynamically using Dirac constraint framework is very easy. The Lagrangian (\ref{a1}) in polar coordinates reduces to
	\begin{equation}
	L=\frac{1}{2\theta}\rho^2\dot{\phi} 
	\label{a4}
	\end{equation}
	that has a pair of Second Class constraints 
	\begin{equation}
	\psi_1\equiv p_\rho,~\psi_2\equiv p_\phi -\frac{1}{2\theta}\rho^2 .	
	\label{a5}
	\end{equation}
	The Dirac brackets are given by
	\begin{equation}
	~	 \{\rho,\phi\}=-\frac{\theta}{\rho},~	 \{\phi,p_\phi\}=1,$$$$ \{\rho,p_\phi\}=0,~\{\phi,p_\rho\}=\{p_\phi,p_\rho\}=	\{\rho,p_\rho\}=0 .		
	\label{a6}
	\end{equation}
	Notice that the NC brackets are operatorial so that one is no longer allowed to take, for instance, a structure 
	$$\{\rho,\phi\}=-\frac{\theta}{\rho},~	\{\rho,p_\rho\}= \{\phi,p_\phi\}=1 $$ with rest of the brackets vanishing since the Jacobi identity 
		$$\{\{\phi,\rho\},p_\rho\}+\{\{\rho,p_\rho\},\phi\}+\{\{p_\rho,\phi\},\rho\}=-\frac{\theta}{\rho^2}\neq 0$$
		is non-zero and the Dirac bracket structure with $\{\rho,p_\rho\}=0$ is demanded. However the Darboux map that we have exploited is difficult to construct (in closed form) for this set operatorial Dirac brackets.


\begin{thebibliography}{99}
		
		
		\bibitem{mal} J. Maldacena, S. H. Shenker, and D. Stanford, JHEP 08,
		106 (2016), 1503.01409.
		\bibitem{mor}T. Morita , Phys. Rev. Lett. 122, 101603 (2019).
		
		\bibitem{mor1}T. Morita (2018), (arxiv:1801.00967).
		
		\bibitem{ncrev}M.R.Douglas and N.A.Nekrasov, Rev. Mod. Phys. 73 977 (2001) [hep-th/0106048];
		R. J. Szabo, Phys. Rep. 378 207 (2003) [hep-th/0109162]; 		R. Banerjee, B. Chakraborty , S. Ghosh , P. Mukherjee , S. Samanta, Found.Phys. 39 (2009) 1297-1345.
		 
		\bibitem{sg1}S. Ghosh, Mod.Phys.Lett. A20 (2005) 1227-1238.
		\bibitem{pai}	A. Pais and G.E. Uhlenbeck, Phys. Rev. 79, 145 (1950).
		\bibitem{pais} A. V. Smilga, Phys. Lett. B 632, 433 (2006) [hep-th/0503213]; R. Banerjee, “New (Ghost-Free) Formulation of the Pais-Uhlenbeck Oscillator,”
		arXiv:1308.4854 [hep-th];  S. Pramanik and S. Ghosh, Mod. Phys. Lett. A 28, 1350038 (2013)
		 [arXiv:1205.3333 [math-ph]]; P.  Matej ,International Journal of Geometric Methods in Modern Physics, Vol. 13, No. 09, 1630015 (2016),
		(arXiv:1607.06589) [gr-qc].
		\bibitem{complexli}S. Wieczorek, {\it Lyapunov Exponents}, 		 
		Applied Mathematics, University College Cork
		May 11, 2016.
	\bibitem{ncpolar}J. P. Edwards,	Eur. Phys. J. C (2017) 77:320.
	
			
	\bibitem{dir}P. A. M.Dirac, Lectures on Quantum Mechanics, Yeshiva University Press, New York,
	1964.
	 \bibitem{sgfluid} 	P. Das, S. Ghosh, Eur.Phys.J. C76 (2016) no.11, 627, Erratum: Eur.Phys.J. C77 (2017) no.2, 64;  
		 P. Das, S. Ghosh,  Phys.Rev. D98 (2018) no.8, 084047.
	 \bibitem{ncfl}  A. K. Mitra, R. Banerjee, S. Ghosh,  JCAP 1810 (2018) no.10, 057.
	\end{thebibliography}
\end{document}